\documentclass{article}

\usepackage{arxiv}

\usepackage[utf8]{inputenc} % allow utf-8 input
\usepackage[T1]{fontenc}    % use 8-bit T1 fonts
\usepackage{hyperref}       % hyperlinks
\usepackage{url}            % simple URL typesetting
\usepackage{booktabs}       % professional-quality tables
\usepackage{amsfonts}       % blackboard math symbols
\usepackage{nicefrac}       % compact symbols for 1/2, etc.
\usepackage{microtype}      % microtypography
\usepackage{lipsum}
\usepackage{graphicx}
\graphicspath{ {./images/} }
\usepackage[T1]{fontenc}
\usepackage{graphicx}  % For including images
\usepackage{subcaption} % For subfigures
\usepackage{svg}
\usepackage{float}
\usepackage{braket}
\usepackage[section]{placeins}
\usepackage{amsmath}
\usepackage{setspace}
\usepackage{caption}
\usepackage{xcolor}
\usepackage{microtype}

\title{Exploring Quantum Heider Balance Theory}

\author{
 Anahid Kiani \\
  Department of Physics, \\
  Shahid Beheshti University,\\
   Evin, Tehran 1983969411, Iran\\
   \And
 S.Mahdi Fazeli \\
  Department of Physics, \\
  University of Qom,\\
   Ghadir Blvd, Qom 3716146611, Iran \\
  \And
 G.Reza Jafari \\
  Chandigarh Group of Colleges Jhanjeri, \\
  Sahibzada Ajit Singh Nagar, Punjab, India.\&
  Department of Physics, Shahid Beheshti University,\\
   Evin, Tehran 1983969411, Iran\\
  \texttt{gjafari@gmail.com} \\
}

\begin{document}
\maketitle

\begin{abstract}
Classical Heider balance theory models the evolution of social networks towards balanced states with stress minimization. Triad relationships are classically either balanced or imbalanced. However, real-world relationships often exhibit uncertainty, complexity, and interconnected dynamics that transcend this classical framework, and we will sometimes see the synchronicity of balance and imbalance states. When these triadics are simultaneously balanced and imbalanced, they form superposition and entanglement states that necessitate the introduction of quantum balance. This study introduces a framework that extends classical Heider balance theory from social network analysis by incorporating principles of quantum mechanics. In the framework, each triad is a quantum state of spin systems embedded in various network topologies. Using a quantum transformation operator, we investigate the emergence of balanced and imbalanced states and identify ground and steady states. In the quantum state, developments towards balance will occur by introducing the Hamiltonian within the creation and annihilation operators framework. In these developments, we witness the emergence of superposition and entanglement quantum states with no classical equivalent, which form an uncertainty in the states. When the quantum evolution is a functional temperature, we investigate the emergence of balanced and imbalanced states, analyze ground and steady states, and examine phase transitions as a function of temperature. Our results reveal novel dynamical behaviors unique to quantum social systems and offer a new perspective on collective decision-making, conflict resolution, and emergent order in complex networks.

\end{abstract}

\flushbottom
\maketitle
\thispagestyle{empty}

%\noindent Please note: Abbreviations should be introduced at the first mention in the main text – no abbreviations lists. Suggested structure of main text (not enforced) is provided below.

\section*{Introduction}

The concept of balance theory, initially introduced by Heider in 1946, has served as a foundational framework to comprehend the social dynamics within networks. This theory posits that individuals seek to preserve cognitive stability within three-way relationships, culminating in balanced or imbalanced configurations \cite{Heiderbalance}.
Over the years, balance theory has been successfully applied to many problems. Cartwright and Harry \cite{cartwright1956structural} built upon Heider's ideas by formalizing the mathematical underpinnings of balance theory, paving the way for the modeling of structural dynamics networks \cite{PRLMarvel2009, Oloomi2021, ctimemarvel, Grski2020, Szell2010, Belaza2017}. Empirical studies, such as those by Antal \cite{ANTAL2006130}, Karpinski \cite{Krapivsky}, and Redner \cite{Render}, have demonstrated the influence of balance dynamics on the evolution of network structures, showing that imbalanced configurations tend to evolve into balanced states over time. The applications of balance theory are numerous, encompassing domains such as political polarization \cite{Talaga2023}, neuroscience \cite{Moradimanesh, Saberi2021}, biology \cite{Rizi2020StabilityOI}, finance \cite{Zahedian2022}, social science \cite{kleinberg, Glassy, Murase2019, Estrada2014, pham2020}, and the evolution of structural networks \cite{Sreenivasan2016, Sibon2025}. 

Classical balance theory has three properties that need to be revised to fit the real world: a) First, each triad is strictly labeled as balanced or imbalanced, with no room for ambiguity; b) transitions between states are discrete and localized, typically involving a single-link modification. c) Triads are treated as independent entities, ignoring the influence of a broader network context.  These assumptions conflict with empirical evidence showing that social relationships are evolving, with some uncertainty, and are strongly interconnected \cite{Biamonte2019, Perseguers2010, Chepuri2023}. In other words, a triad relationship has its benefits (balance) or stresses (imbalance) at the same time.
We propose a quantum extension of Heider's balance theory to address these limitations. By representing each triad as a quantum state, we incorporate the principles of superposition and entanglement, capturing the probabilistic, nonlocal, and interdependent nature of human interactions. This quantum formalism allows for more realistic modeling of social systems, where a trial relationship may simultaneously embody trust and doubt, and where the balanced state of one triad can influence others. The framework underscores the advantages of quantum approaches in comparison to classical models and initiates new directions for various research projects in quantum cognition and decision-making \cite{maksymov2024quantum6, Bagarello_2017}, biology \cite{McFadden1999AdaptiveMutation5}, neuroscience \cite{Khrennikov2018QuantumDecisionMaking, BASIEVA2024}. 

Contrary to classical balance theory, a) In real-world social systems, triadic relationships are seldom purely balanced or imbalanced; instead, they are often marked by inherent uncertainty. This uncertainty reflects the ambiguity and complexity of interpersonal dynamics, where relationships may harbor doubt, tension, or conflicting tendencies. As a result, larger network structures emerging from these foundational triads cannot be simply classified as positive or negative. Instead, each triad exists in a state that eludes binary categorization, better conceptualized as a superposition of balanced and imbalanced configurations. The realization of a positive (balance) or negative (imbalance) outcome depends on the dynamics of interaction, underscoring the fundamentally probabilistic nature of social balance. b) When each triad is described as a superposition of positive and negative states, transitions within the network no longer follow discrete shifts between balanced and imbalanced configurations. Instead, the system exhibits continuous transformations, reflecting the nuanced and fluid nature of social dynamics. In reality, human relationships cannot be reduced to tunable parameters that yield balance through simple adjustments of individual links. Rather, relational change tends to unfold gradually, aligning more closely with a continuous formalism that captures the evolving and often ambiguous character of interpersonal interactions. c) Moreover, in real-world social systems, the imbalance within one triad can influence and even destabilize adjacent balanced configurations. This interplay resonates with the principles of quantum entanglement, wherein the state of one entity nonlocally affects the state of another, highlighting the interconnected and context-dependent nature of relational dynamics. 

In this work, we model social triads as spin systems embedded in various network topologies. Using tools from quantum mechanics, we construct a Hamiltonian that governs the dynamics of balance and imbalance within the network. We analyze how quantum operators drive the system's evolution, study the spectral properties of the Hamiltonian, and explore phase transitions under finite-temperature conditions. Our results highlight the emergence of novel behaviors, such as coherent transitions and entangled dynamics, not captured by classical models. This quantum framework opens a new avenue for studying collective behavior, conflict resolution, and systemic change in complex social networks. propelling evolutionary optimization for intricate systems \cite{DENG20241,PhysRevResear4}.

\section{Classical Heider balance theory}
The theory of balance is based on triadic groups of relationships in which every two individuals have a relationship of enmity (-1) or a relationship of friendship (+1).  In the case of three-body interactions, four different configurations arise due to the two possible signs for each pairwise relation. The (+ + +) , (+ - -) and its permutations, (+ + -) and its permutations, and (- - -). Triads with an even (odd) number of negative relationships are called balanced (imbalanced). The primary objective of balance theory is to reduce tension and the number of imbalanced triads within a network of relations, thereby facilitating the transition towards a balanced state.  
The transformation model of marked networks was first explored by Antal et al. \cite{ANTAL2006130}. The primary objective of balance theory is to reduce tension and the number of imbalanced triads within a network of relations, thereby facilitating the transition towards a balanced state. Two models were proposed to guide an imbalanced network to equilibrium: local triad dynamics (LTD) and the constrained triad dynamics (CTD). Our approach is based on the LTD model. At each time step of the triad, a random selection is made from the set of all potential dynamic models of the local triad. If the triad is balanced, no changes will be made at the zero temperature. However, it will be changed by the Boltzmann distribution function at nonzero temperatures. 
Over the years, many researchers have contributed to the development of the balance theory. Also, many applications in various fields of neuroscience, biology, sociology, and politics have been reported using this theory.
In \cite{ctimemarvel}, the authors put forth the discrete time and continuous time models, respectively, as a means of studying the dynamics of balance theory on a fully connected network. The dynamics of a network comprising three-body interactions in a nonzero temperature environment have been the subject of investigation. The spectrum of energy in balance theory can be defined under the definition of balance states proposed by Marvel, which formed the foundation for the procedure utilized to develop balance theory.
In \cite{Krapivsky}, the authors propose the discrete time and continuous time models, respectively, to study the dynamics of balance theory on a fully connected network. The dynamics of a network of three-body interactions in a nonzero-temperature environment have been investigated.
Bagherikalhor et al in \cite{Bagherikalhor} employ the balance theory in a fully connected network with weighted triads, where the weights are drawn from a normal distribution. Marvel et al. \cite{PRLMarvel2009} define the energy landscape for networks, demonstrating that networks may be trapped in local minima, or jammed states, which depend on the size of the network. Belaza et al \cite{Belaza} investigate balance theory from the statistical physics point of view and model the triadic energy, testing it on a dataset of relations between countries during the Cold War era. Balance theory has also been applied to brain networks to distinguish individuals with autism spectrum disorder from healthy controls \cite{Moradimanesh}. The spectrum of energy in balance theory can be defined by reference to the definition of balance states proposed \cite{ctimemarvel}, which formed the basis of the procedure used to develop balance theory:
\begin{equation}
    H = - \sum_{i>j>k}^{n} s_{ij}s_{jk}s_{ki}.
\end{equation}
$s_{ij}$ is the edge between the $i$th node and the $j$th node in the network.
Following this, numerous studies have been conducted that examine the dynamics of networks by considering triadic interactions.
The statistical properties of the system were calculated using Hamilton’s approach to statistics, in conjunction with the Boltzmann distribution function and the mean-field method. This enabled the study of the system's behavior and the relationship between its various components.
\begin{figure}
    \centering
    \includegraphics[width=0.9\linewidth]{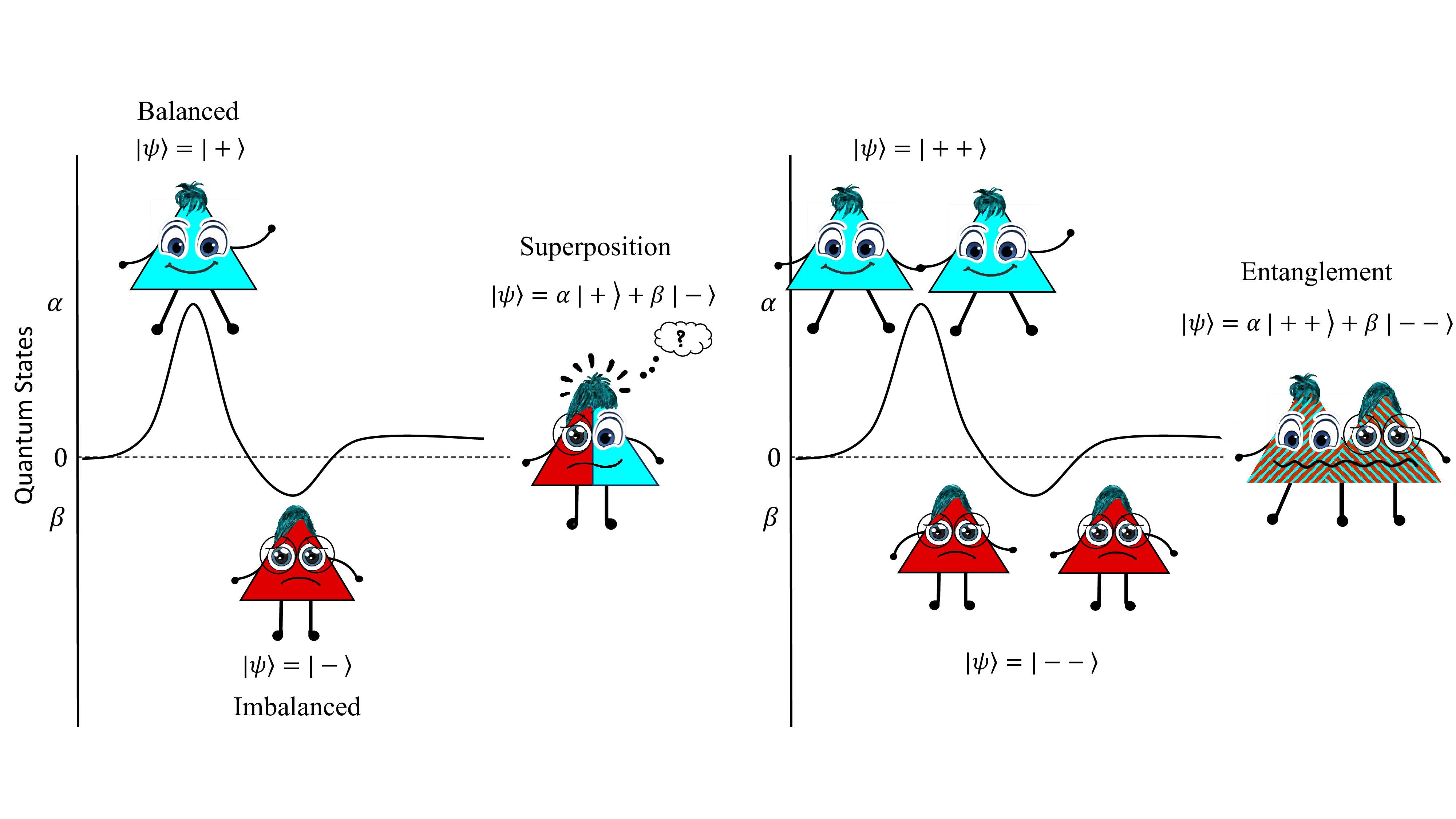}
    \caption{Schematic representation of balance and imbalance quantum triplet states. The left panel depicts the formation of a superposition state, and the right panel depicts an entangled state.}
    \label{fig1}
\end{figure}

\section{Mathematical Formalism of Quantum Balance Theory} 
The classical Heider balance theory is predicated on the categorization of triadic relations as balanced or imbalanced, relying on binary labels that overlook the subtleties of real-world social interactions. Human relationships are frequently characterized by fluidity, uncertainty, and interconnectedness, qualities that are not adequately captured by classical formulations. To address these limitations, a quantum extension of balance theory was developed, incorporating the principles of superposition, entanglement, and quantum dynamics.
In what follows, we construct a rigorous mathematical framework for quantum balance theory by integrating concepts from quantum mechanics and network science. We begin by formally defining the quantum states representing social triads, then introduce the transformation operators that drive state transitions. We proceed to build the system's Hamiltonian and analyze its spectral properties to predict network dynamics.

\subsection{Quantum Representation of Triad States}
\begin{figure}[t]
   \centering\includegraphics[width=1\linewidth]{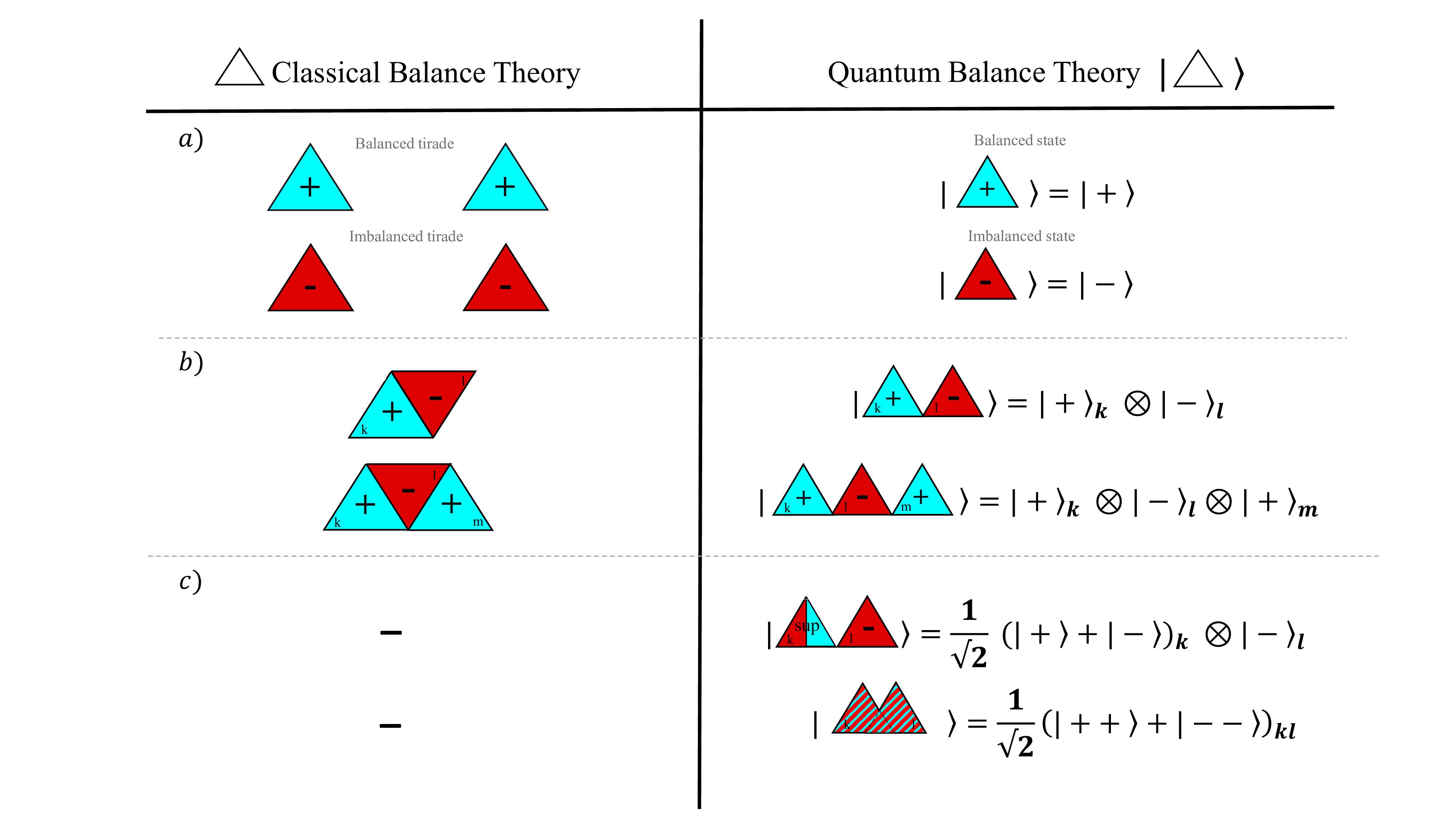}
	\caption{(a) Classical triads (left) are shown alongside their quantum analogues (right), where each triad is encoded as a quantum state.
(b) The system's states (comprised of two and three triads in this specific representation) arising from tensor products of single-qubits.
(c) Superposition
$\ket{\psi_{Sup}} = \frac{1}{\sqrt{2}}(\ket{+-}+\ket{--})$ and entanglement $\ket{\psi_{Ent}} = \frac{1}{\sqrt{2}}(\ket{++}+\ket{--})$ visually demonstrated here (right) have no classical counterparts.} {\label{fig2}}
\end{figure}

We assign a quantum state to each triad to investigate the connection between quantum mechanics and classical balance theory, thereby embedding relational configurations within a Hilbert space. Specifically, we associate the spin-up state $\ket{+}$ with a balanced triad and the spin-down state $\ket{-}$ with an imbalanced one. Fig. \ref{fig1} schematically shows quantum triad superposition and entangled states. Each of the triads is thus modeled as a quantum state $\ket{\Delta}$ in a two-dimensional Hilbert space spanned by the basis $\{\ket{+},\ket{-}\}$ For a network of $N$ such triads, the tensor product gives the state of the system:
\begin{equation}
    \ket{\psi(t)} = \bigotimes_{n=1}^{N} |\Delta_n\rangle.
\end{equation}
This formalism captures the possibility of each triad existing in a coherent superposition of balanced and imbalanced states, thereby reflecting the probabilistic and uncertain nature of human beings. 
As illustrated in Fig. \ref{fig2}, we present a comparative analysis of triads in classical and quantum balance theory. In Fig. \ref{fig2}a, the left column depicts the classical triads, while the corresponding quantum analogues are shown in the right column. Fig. \ref{fig2}b extends this comparison by presenting a broader set of triadic configurations that give rise to quantum states via tensor products. Notably, Fig. \ref{fig2}c highlights the emergence of quantum features—superposition and entanglement—for which no classical equivalents exist.
The evolution of the system is directed toward reducing imbalance and promoting stable, balanced configurations. This process is governed by a Hamiltonian operator that encodes the transition dynamics between states and determines the time evolution of the network. The influence of quantum properties such as superposition and entanglement is incorporated through spectral decomposition of the Hamiltonian, enabling a detailed analysis of how initial configurations evolve. This quantum perspective offers a distinctive framework for capturing the inherently probabilistic and complex nature of social dynamics.
This formalism allows us to investigate how balanced and imbalanced configurations emerge and evolve within a quantum network. To illustrate the approach, we present a minimal example of two interconnected triads, whose states can be categorized into three distinct classes: separable, superposed, and entangled.

\subsection{Single-qubit states}
As illustrated in Fig. \ref{fig2}a, each triad is described by a single qubit, balanced, and imbalanced state. The matrix representation of this quantum state is given by:
\begin{align}
  \ket{\Delta_+}=\ket{+} = \begin{pmatrix} 
	1\\
	0\\
    \end{pmatrix},     \ \ \ \ \ \ \ \ \ \  \ \ \ \  \  \  \ \
    \ket{\Delta_-} = \ket{-} = \begin{pmatrix} 
        	0\\
        	1\\
        \end{pmatrix}.  
\end{align}
Considering a quantum system consisting of two triads, denoted by n = k and n = l, as shown in Fig. \ref{fig2}b, the basis states for this system can be defined as the tensor product of the individual basis states, as follows:
$\ket{+-}$, $\ket{-+}$, $\ket{--}$ and $\ket{++}$. 
%Expanding this framework to three qubits, designated as n = k, l, and m, the system’s basis states form via the tensor product of each triad’s basis states: $ \ket{---} $, $ \ket{--+} $, $ \ket{-+-} $, $\dot$ $\dot$ $ \dot$, $\ket{+++} $.
Expanding this framework to three qubits, designated as $n = k, l, \text{and } m$, the system’s basis states form via the tensor product of each triad’s basis states: $\ket{---}$, $\ket{--+}$, $\ket{-+-}$, $\ldots$, $\ket{+++}$.

%In the absence of entanglement among these three quibits, their states remain independent, permitting the expression of the overall system’s state as:??. 
In light of network dynamics, particularly within the context of balance theory, our primary objective is to achieve equilibrium and, consequently, to maximize the prevalence of balanced triads, which represent a state of optimal stability.        
%\begin{equation*}
   %\ket{\Delta_+ \ \Delta_+} = \ket{+} \otimes \ket{+}, \\\ket{\Delta_- \ \Delta_+} = \ket{-} \otimes \ket{+}, \\\ket{\Delta_+ \ \Delta_-} = \ket{+} \otimes \ket{-}, \\\ket{\Delta_- \ \Delta_-} = \ket{-} \otimes \ket{-} \end{equation*}
\subsection{Quantum Superposition and the Probabilistic Nature of Triads}
Classical balance theory categorizes triadic relations as either balanced or imbalanced, based on deterministic rules. Yet, in real-world social contexts, individuals often experience conflicting emotions or mixed sentiments simultaneously \cite{haven2013} conditions that challenge this binary classification. Quantum superposition offers a compelling mathematical foundation for modeling such uncertainty, enabling a triad to exist in a coherent mixture of balanced and imbalanced states rather than a fixed configuration. This perspective aligns with findings in psychology and behavioral economics, where human decision-making frequently exhibits probabilistic and context-dependent characteristics—features well described by quantum probability theory \cite{Bruza2012}. In this framework, as depicted in Fig. \ref{fig2}c, each triad is represented as a quantum superposition of balanced and imbalanced states, a representation that cannot be captured within classical models. As illustrative examples, we consider specific superposition states constructed from pairs of triads, showcasing the expanded expressive capacity of the quantum formalism.
\begin{align}
   \ket{\Delta_{Sup} \  \Delta_{\pm}} = \frac{1}{\sqrt{2}}(\ket{+} \pm \ket{-}) \otimes \ket{\pm} , \ \ \ \ \ \ \ \ \ \  \ \ \ \  \  \  \ \ 
   \ket{\Delta_{Sup} \  \Delta_{Sup}} =  \frac{1}{2}(\ket{+} \pm \ket{-}) \otimes (\ket{+} \pm \ket{-}).
\end{align}
\subsection{Entanglement and the Nonlocal Coupling of Triads}
Classical balance theory treats each triad as an isolated unit, evolving independently from others. In contrast, real-world social relationships are often interdependent, with changes in one triad influencing others through shared individuals or collective dynamics. Within the quantum formalism, this interdependence is naturally captured by the concept of entanglement.
In an entangled state, the configuration of one triad cannot be described independently of another. Instead, the joint state encodes correlations that extend nonlocally across the network. Empirical support for this idea emerges from cognitive science, where entanglement-like patterns have been observed in perception and decision-making processes \cite{asano2015sound}. Similarly, recent developments in quantum cognition suggest that non-classical dependencies in human reasoning can be effectively modeled using quantum correlations \cite{Pothos2013}.
In this regime, as depicted in Fig. \ref{fig2}c, the collective state of two entangled triads is represented by a single, inseparable quantum state, underscoring the deeply intertwined nature of social dynamics. We illustrate this concept with explicit examples of entangled states shared between two triads, highlighting the departure from classical, locally defined configurations.
\begin{align}
   \ket{\Delta \  \Delta}_{Ent} =  \frac{1}{\sqrt{2}}(\ket{++} \pm \ket{--}) ,\ \ \ \ \ \ \ \ \ \  \ \ \ \  \  \  \ \
   \ket{\Delta \  \Delta}_{Ent} =  \frac{1}{\sqrt{2}}(\ket{+-} \pm \ket{-+}).
\end{align} 
In quantum balance theory, the transformation of imbalanced triads into balanced ones is not an isolated process but is influenced by shared links and interdependencies within the network. As a result, the evolution of one triad can affect others, leading to dynamic, system-wide transitions. These intertwined quantum states reflect the nonlocal nature of social influence. The system evolves through changes in its collective wave function, governed by a Hamiltonian operator. To construct this Hamiltonian, we first introduce the necessary quantum operators using the formalism of second quantization.

\begin{figure}[t]
	\centering\includegraphics[width=1.0\linewidth]{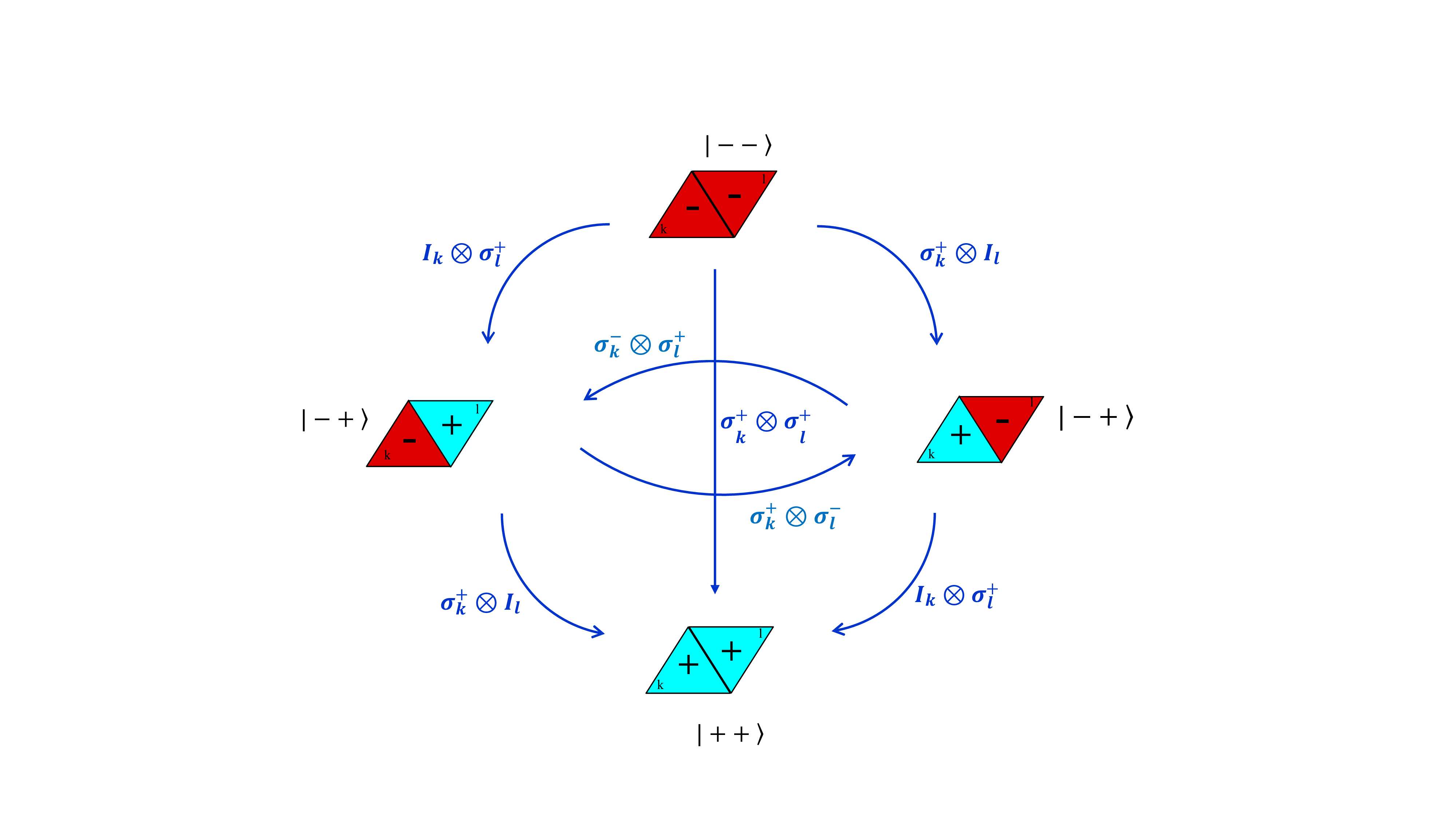}
	\caption{Transition pathways between balanced and imbalanced states in a two-triad quantum system: 
The diagram illustrates all possible transitions among the basis states of a two-triad network, driven by the application of creation $\sigma^{+}$ and annihilation $\sigma^{-}$ operators. Each link represents an allowed quantum transition between balanced and imbalanced configurations of the triads. The transitions encode the system’s probabilistic dynamics toward reducing imbalance and form the foundation for constructing the network Hamiltonian that governs social evolution in the quantum balance framework.}{\label{fig3}}
\end{figure}

\subsection{Creation and Annihilation Operators}
To model the dynamic evolution of balance in quantum networks, we introduce transition operators that act analogously to ladder operators in quantum mechanics. These operators govern transformations between the balanced $\ket{+}$ and imbalanced $\ket{-}$ states of triads, enabling the system to evolve toward configurations of lower tension.
We define the creation and annihilation operators as:
\begin{align}
    \sigma^+ = \ket{+} \bra{-} = \begin{pmatrix} 
		0 & 1 \\
            0 & 0
		\end{pmatrix},  \ \ \ \ \ \ \ \ 
    \sigma^- = \ket{-} \bra{+} = \begin{pmatrix} 
		0 & 0 \\
            1 & 0
		\end{pmatrix}.
\end{align}
These operators perform elementary transitions: $\sigma^{+}$ maps an imbalanced triad to a balanced one, while $\sigma
^{-}$ performs the reverse. Their action is constrained by the boundary conditions of the system:
\begin{equation}
    \sigma^+ \ket{-} = \ket{+} ,\\ \sigma^+ \ket{+} = 0,\\
   \sigma^- \ket{+} = \ket{-}  ,\\  \sigma^- \ket{-} = 0.
\end{equation}
Unlike the unbounded ladder operators of the harmonic oscillator, these operators act within a two-level system with clearly defined upper and lower states.
To quantify the balance state of a given triad, we introduce the occupation number operator $\hat{n}$ \cite{Derrida1993}: 
\begin{align}
     \hat{n} = (\frac{1}
  {2})\begin{pmatrix} 
		1 & 0 \\
            0 & -1
		\end{pmatrix}, \ \ \  \ \ \ \hat{n} \ket{\pm} = \pm\frac{1}{2}\ket{\pm}.
\end{align}
Its action on a given state measures the value that is currently stored in the state.
\section{Formulation of the Transition Hamiltonian in Quantum Dynamics}
The evolution of quantum networks is governed by a Hamiltonian that encodes probabilistic transitions between balanced and imbalanced triadic configurations. We employ the creation and annihilation operators introduced earlier to construct this operator, which mediates state transitions within and between social triads.
We begin by formulating the Hamiltonian for a minimal system composed of two interacting triads, each represented as a two-level quantum system. In this setting, the transition from imbalance to balance—and vice versa—is captured by applying ladder operators that preserve the inherent uncertainty and interdependence of quantum social states. All of the possible transitions within this two-triad system is illustrated in Fig. \ref{fig3}, where each pathway corresponds to an operator-induced transformation between states. The Hamiltonian for this two-triad configuration takes the form:
\begin{equation}
     H = \sum_{k>l=1}^{N=2} [ \sigma^+_k \otimes \mathbb{I}_l +\sigma^+_k \otimes \sigma^-_l + \sigma^+_k \otimes \sigma^+_l + diagonal \ sentences],
\end{equation}
where $\sigma^{+}$ and $\sigma^{-}$ denote the ladder operators acting on the $k$-th and $l$-th triads, respectively, and $\mathbb{I}$ is the identity operator. 
%The non-diagonal terms represent state transitions, while the diagonal terms account for self-interactions and preserve normalization. This pairwise construction forms the basis for generalizing the Hamiltonian to networks of arbitrary size. In the extended framework, the Hamiltonian incorporates both local transformations and triad-level interactions, enabling a scalable representation of quantum social evolution. 
%This formalism allows for the investigation of emergent behaviors, including convergence toward balance and the propagation of imbalance across entangled social structures.
Non-diagonal elements of the Hamiltonian encode transitions between distinct quantum states, enabling the system to evolve dynamically over time. In contrast, diagonal elements represent self-interactions, indicating the likelihood of a state remaining unchanged. To ensure conservation within the system’s probabilistic evolution, the Hamiltonian is constructed such that the sum of each column in its matrix representation vanishes. This condition guarantees that the total probability flux remains balanced, reflecting the closed and norm-preserving nature of quantum state evolution.
\subsection{Probability Conservation and Generalized Hamiltonian Structure}
To ensure consistency with quantum probability conservation, we define the vector $\bra{S}$ as the linear superposition of all possible basis states: 
\begin{align}
    \bra{S} = \bra{+} + \bra{-}.
\end{align}
This vector acts uniformly across the entire state space, serving as a test for the conservation of total probability under the Hamiltonian dynamics. For the system to evolve in a physically consistent manner, the Hamiltonian $H$ must satisfy the condition:
\begin{align}\label{eq_16}
    \bra{S} H = 0.
\end{align}
This requirement ensures that each state's net probability flow into and out is balanced, with no artificial gain or loss across the full system. In matrix language, it demands that the sum of each column of $H$ vanishes, a standard property of stochastic or norm-preserving quantum operators.
The action of the transition operators on $\bra{S}$  can be explicitly evaluated. Applying the creation operator yields:
\begin{align}
    \bra{S} \sigma^+ = \bra{+} \sigma^+ + \bra{-}  \sigma^+ = (\sigma^- \ket{+})^{\dagger} + (\sigma^- \ket{-})^{\dagger} = \bra{-} + 0\bra{+}  = - \bra{S}(\hat{n} - \frac{1}{2} \mathbb{I}), 
\end{align}
where $\hat{n}$ is the operator of the occupation number and $\mathbb{I}$ denotes the identity.
Similarly, applying the annihilation operator gives:
\begin{align}
    \bra{S} \sigma^- = \bra{+} \sigma^- + \bra{-}  \sigma^- = (\sigma^+ \ket{+})^{\dagger} + (\sigma^+ \ket{-})^{\dagger} = 0\bra{-} + \bra{+} = \bra{S}(\hat{n} + \frac{1}{2} \mathbb{I}). 
\end{align}
These relations are crucial for verifying the conservation condition $\bra{S} H = 0$ once the full Hamiltonian is assembled.
Extending this framework to networks of arbitrary size, the general Hamiltonian for a system of $N$ triads is constructed as 
\begin{align}{\label{eq_H}}
    H =& \sum_{k>l=1}^N  [\sigma^+_k 
    -(\frac{1}{2} \mathbb{I}_k - \hat{n}_k)]\otimes \mathbb{I}_l
    + \sum_{k>l=1}^N [\sigma^+_k \otimes \sigma^-_l
    -(\frac{1}{2} \mathbb{I}_k - \hat{n}_k) \otimes (\frac{1}{2} \mathbb{I}_l + \hat{n}_l) ]
    \nonumber \\ & + \sum_{k>l=1}^N [\sigma^+_k \otimes \sigma^+_l  - (\frac{1}{2} \mathbb{I}_k - \hat{n}_k) \otimes (\frac{1}{2} \mathbb{I}_l - \hat{n}_l)].
\end{align}
This Hamiltonian incorporates local transitions, mutual interactions between triads, and stabilizing terms to ensure conservation of the system's total probabilistic weight. It provides the foundation for analyzing the quantum dynamics of larger social networks under the principles of quantum balance theory.
\section{Spectral Properties and Time Evolution in Quantum Balance Networks}
The dynamical evolution of a quantum social network is governed by the spectral properties of its Hamiltonian \cite{lewin2024spectral}. Decomposing the system into its eigenmodes makes it possible to describe the full time-dependent behavior of any initial state in terms of a superposition of stationary solutions. For a general quantum network governed by Hamiltonian $H$, the state of the system at time $t$ can be expressed as:
\begin{equation}
    \ket{\psi(t)} = \sum_{\lambda} c_\lambda(t)
    \ket{\lambda} ,\\\\\
    c_\lambda(t) = c_\lambda(0) e^{\lambda t}
    ,\\\\\
    \sum_{\lambda} |c_\lambda(t)|^2 = 1,
\end{equation}
where $\ket{\lambda_i}$ and $\lambda_i$ denote the eigenvectors and eigenvalues of $H$, respectively. The coefficients 
$c_{\lambda}(0)$are determined by projecting the initial configuration $\ket{\psi(t=0)}$ onto the eigenbasis:
\begin{align}
    c_{\lambda}(0) = \braket{\lambda|\psi(t=0)} .  
\end{align}
Each eigenvalue governs the temporal evolution of its associated eigenstate, with negative eigenvalues leading to exponential decay and zero eigenvalues indicating steady-state components.
The probability of observing the system in a particular basis state $\ket{i}$ at time $t$ is given by:
\begin{align}{\label{eq_p}}
    P_i(t) = |\braket{i|\psi(t)}|^2, 
\end{align}
allowing a full reconstruction of the evolving probability distribution over all network configurations. This spectral approach enables the characterization of how initial imbalances and superpositions dissipate, how entanglement structures evolve, and how the system asymptotically approaches balance.
\subsection{Application to a Two-Triad System}
To illustrate this formalism, it is applied to a minimal quantum network consisting of two interacting triads. The Hamiltonian for this system is as follows, according to Eq. (\ref{eq_H}), yielding four eigenvalues:
\begin{align}\label{eq22}
     H = \begin{pmatrix} 
		0 & 1 & 1 & 2 \\
            0 & -2 & 1 & 1 \\
            0 & 1 & -2 & 1 \\
            0 & 0 & 0 & -4
		\end{pmatrix}.
\end{align}
Diagonalization of $H$ reveals four eigenvalues:
\begin{align}\label{eq24}
  \lambda_1 =  -4, \   \  \  \  \  \  \  \  \   \  \  \  \ \lambda_2 =  -3, \  \  \  \  \  \  \   \  \  \  \  \  \   \lambda_3 =  -1, \  \  \  \  \  \  \  \  \  \  \  \  \lambda_4 =  0.    
\end{align}
with corresponding eigenvectors:
\begin{figure}[t]	\centering\includegraphics[width=1.0\linewidth]{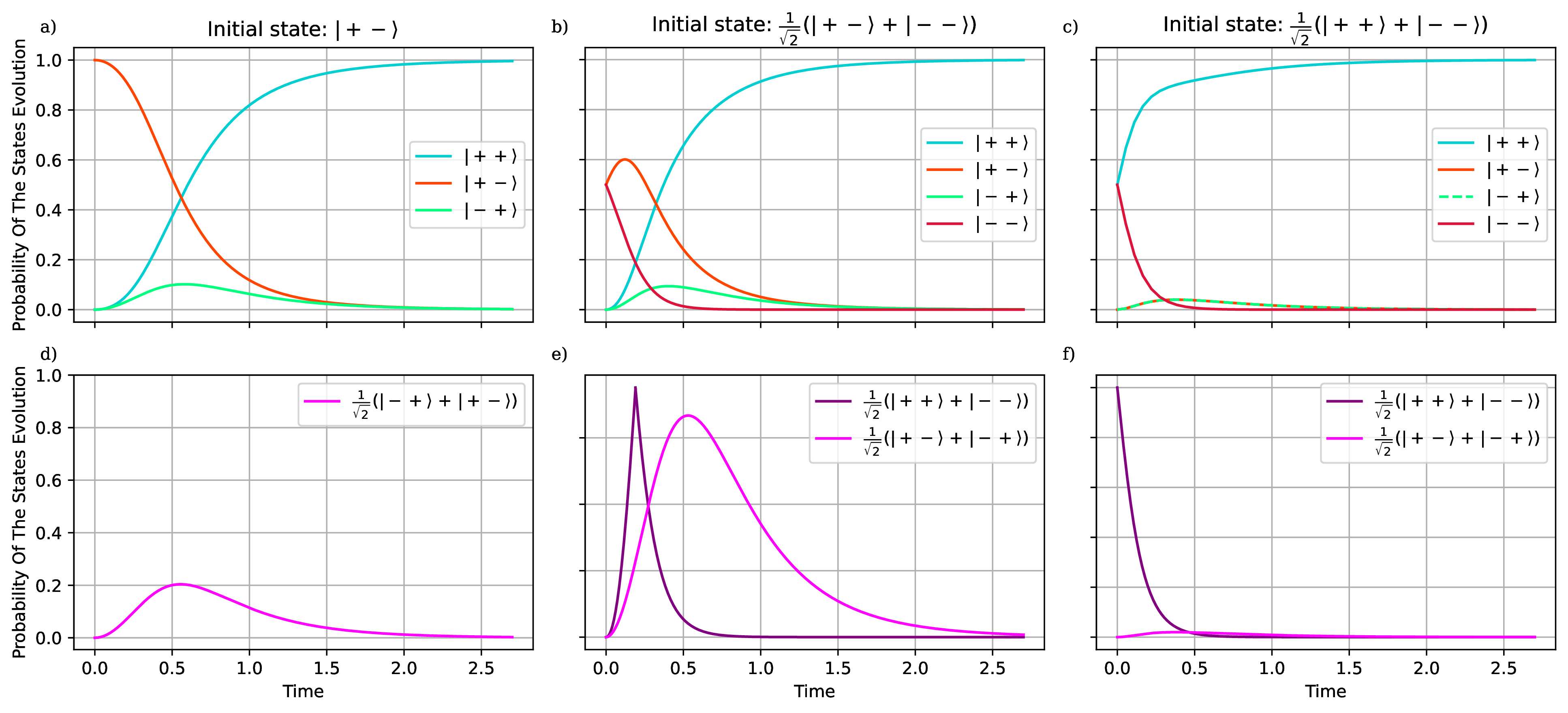}
    \caption{Evolution of quantum state probabilities over time for different initial states. The initial state is represented by $(a) $ $ \ket{+-}$; $(b)$ superposition, $\frac{1}{\sqrt{2}}(\ket{+-} + \ket{--})$; $(c)$ entanglement, $\frac{1}{\sqrt{2}}(\ket{++} + \ket{--})$. The initial states undergo a rapid transition characterized by the decay of certain components, while a particular term,$\ket{++}$, progressively dominates the system's evolution. $(d)$ The entangled state $\frac{1}{\sqrt{2}}(\ket{-+} + \ket{+-})$ emerged during evolution from the initial state $\ket{+-}$ $(e)$ The entangled states $\frac{1}{\sqrt{2}}(\ket{++} + \ket{--})$ and $\frac{1}{\sqrt{2}}(\ket{+-} + \ket{-+})$ also became apparent from the initial state of $(b)$, however, in $(f)$ $\frac{1}{\sqrt{2}}(\ket{++} + \ket{--})$ was present from the starting point, and the entangled state $\frac{1}{\sqrt{2}}(\ket{+-} + \ket{-+})$ arose during evolution.   }{\label{fig4}}
\end{figure}
\begin{align}
    \ket{\lambda_1} = \begin{pmatrix} 
	-1\\
	-1\\
        -1\\
         3
    \end{pmatrix}
       \ , \ket{\lambda_2} = \begin{pmatrix} 
	0\\
	-1\\
        1\\
         0
    \end{pmatrix}
    \ , \ket{\lambda_3} = \begin{pmatrix} 
	-2\\
	1\\
        1\\
         0
    \end{pmatrix}
      \ , \ \ \ket{\lambda_4} =  \begin{pmatrix} 
	1\\
	0\\
        0\\
         0
    \end{pmatrix}.
\end{align}
The eigenvectors can be expressed as a linear combination of the basis states $\ket{++}$, $\ket{+-}$, $\ket{-+}$, and $\ket{--}$: To illustrate this point, we expand the eigenvectors in terms of basis:
\begin{align}
     \ket{\lambda_1} &= -\ket{++} -\ket{+-} -\ket{-+} + 3\ket{--} \\
    \ket{\lambda_2} &= -\ket{+-} + \ket{-+} \\
    \ket{\lambda_3} &= -2\ket{++} +\ket{+-} +\ket{-+}  \\
    \ket{\lambda_4} &= \ket{++}. 
\end{align}
Given an initial state $\ket{\psi(t=0)}$,  the corresponding coefficients $c_{\lambda}(0)$  are calculated by projection onto the eigenvectors.
%The complete time-dependent state is then reconstructed as:
%\begin{equation}
 %   \ket{\psi(t)} = \sum_{i=1}^4 c_{\lambda_i}(0) e^{\lambda_i t}
  %  \ket{\lambda_i} 
%\end{equation}
The initial configuration of the system can correspond to single-qubit triads, coherent superpositions, or entangled states, 
and the corresponding probability distribution across the basis states can be computed.
In this analysis, we consider three distinct initial conditions:
\begin{align}
     \ket{\psi(t_0)} = \ket{+} \otimes \ket{-}=\ket{+-},\ \ \ \ \ \ket{\psi(t_0)} = \frac{1}{\sqrt{2}} (\ket{+} + \ket{-}) \otimes \ket{-}, \ \ \ \  \ \ket{\psi(t_0)} = \frac{1}{\sqrt{2}} (\ket{++} + \ket{--}).
     \label{eq25}
\end{align}
For each case, the probability of finding the system in one of the configurations, such as $\ket{++}$, $\ket{+-}$, $\ket{-+}$ and $\ket{--}$, and so on, is determined according to the projection formula outlined in Eq. (\ref{eq_p}).
As an illustrative example, for the second initial condition: 
$ \ket{\psi(t_0)} = \frac{1}{\sqrt{2}} (\ket{+-} + \ket{--})$, the time-dependent quantum state is as follows:
\begin{align}
\ket{\psi(t)} = \ & 
    \frac{-1}{3\sqrt{2}}(e^{-4t}+5e^{-t}-6) \ \ket{++} \ + \ 
    \frac{1}{\sqrt{2}}\left(\frac{-1}{3}e^{-4t}+\frac{1}{2}e^{-3t}+\frac{5}{6}e^{-t}\right) \ \ket{+-} \nonumber \\
& + \ 
    \frac{1}{\sqrt{2}}\left(\frac{-1}{3}e^{-4t}-\frac{1}{2}e^{-3t}+\frac{5}{6}e^{-t}\right) \ \ket{-+} \ + \ 
    \frac{1}{\sqrt{2}}e^{-4t} \ \ket{--}.
\end{align}
In consideration of the system state function $\ket{\psi(t)}$, it is possible to ascertain the probability of the system being in each of the basic states over time:
  
\begin{align}
   P_{++}(t) &= |\braket{++|\psi(t)} |^2 = |\frac{-1}{3\sqrt{2}}(e^{-4t}+5e^{-t}-6)|^2 \\ 
   P_{+-}(t) &= |\braket{+-|\psi(t)} |^2 = |\frac{1}{\sqrt{2}}(\frac{-1}{3}e^{-4t}+\frac{1}{2}e^{-3t}+\frac{5}{6}e^{-t})|^2 \\
   P_{-+}(t) &= |\braket{+-|\psi(t)} |^2 = |\frac{1}{\sqrt{2}}(\frac{-1}{3}e^{-4t}-\frac{1}{2}e^{-3t}+\frac{5}{6}e^{-t})|^2 \\
   P_{--}(t) &= |\braket{--|\psi(t)} |^2 = |\frac{1}{\sqrt{2}}e^{-4t}|^2. 
\end{align}
\begin{figure}[t]
 \centering\includegraphics[width=1.0\linewidth]{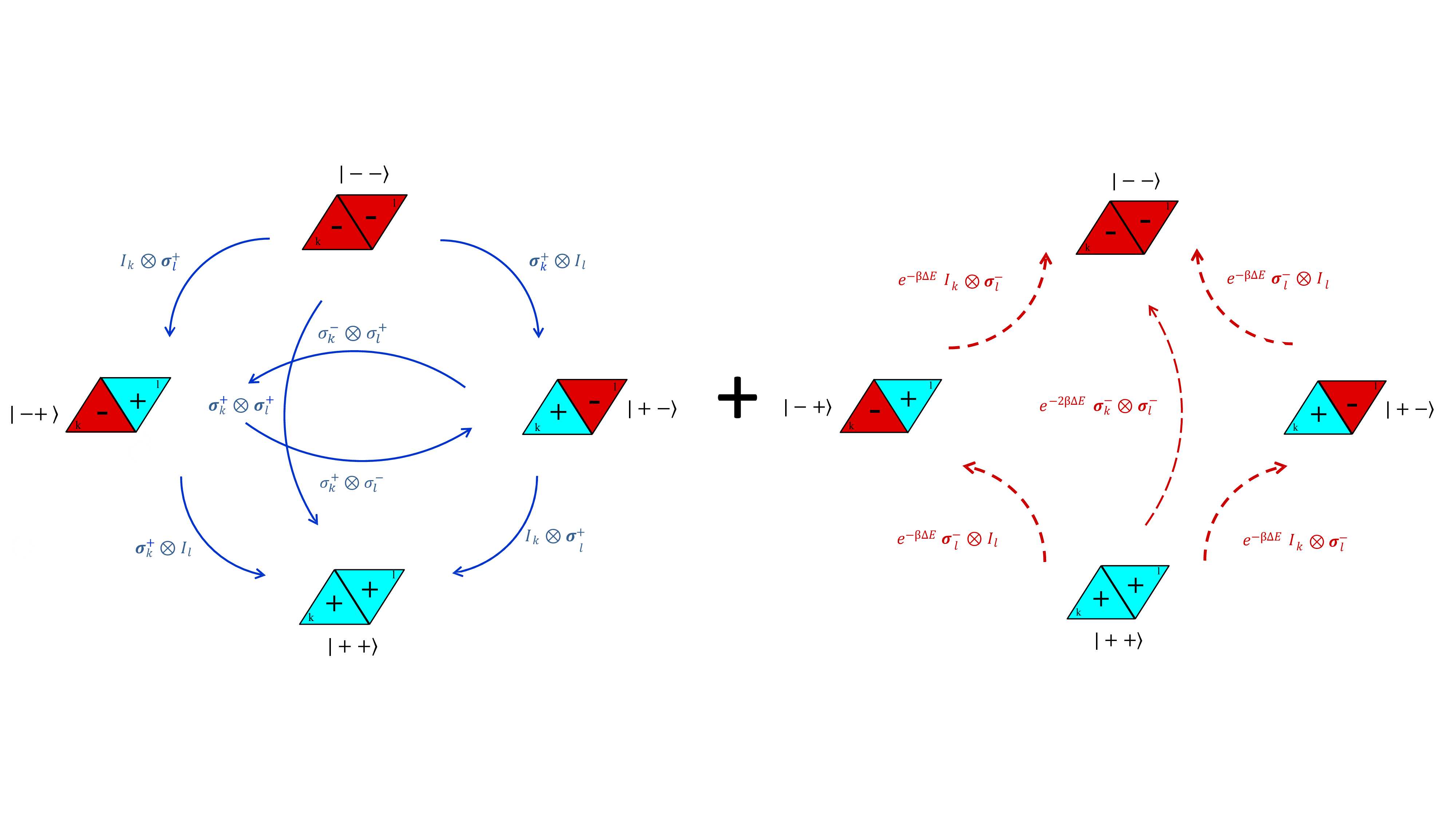}
	\caption{As demonstrated in the accompanying figure, the transition dynamics of the triad between balanced and imbalanced states are illustrated. Each transition represents the application of quantum operators and temperature-dependent factors that govern the system's evolution. In this evolution, the balanced triads are permitted to convert to an imbalanced state with the control parameter, temperature.}{\label{fig5}}
\end{figure}
The initial states derived from Eq.(\ref{eq25}) correspond to those shown in Fig. \ref{fig4}a, Fig. \ref{fig4}b, and Fig. \ref{fig4}c, respectively, where the temporal evolution of the system's probability of remaining in the fundamental states is depicted. The entangled states associated with each of these initial conditions are illustrated in Fig. \ref{fig4}d, Fig. \ref{fig4}e, Fig. \ref{fig4}f. As demonstrated by the diagram, the system is in the initial state at the initial time. It is evident that, over time, the evolution of states progresses in a direction where the probability of the system being in all imbalanced states tends to zero. Consequently, it is the state in which the system is in the most balanced possible state that survives. These results highlight how the system’s evolution toward balance is shaped by its initial quantum structure, with distinct pathways emerging depending on whether the initial state is separable, superposed, or entangled.

\section{Quantum Balance Dynamics at Finite Temperature}
The previous section's analysis assumed an evolution directed toward the reduction of imbalanced triads, corresponding to a zero-temperature limit. However, two important limitations of this approach prompt a broader investigation. First, real systems can undergo transitions that increase the number of imbalance triads, introducing energy into the system. Such processes naturally correspond to finite temperatures. Second, the behavior under nonzero temperatures may qualitatively differ, potentially leading to critical phenomena analogous to those observed in classical balance theory. At nonzero temperatures, the system is allowed to transition toward more balanced configurations and more imbalanced ones, with probabilities weighted by the Boltzmann factor. Specifically, imbalance triads are granted a finite probability to form and persist, proportional to $e^{-\Delta E/T}$, where $e^{-\Delta}$ is the energy cost associated with imbalance, and $T$ is the temperature. This modification raises the central question of whether a phase transition occurs: below a critical temperature, the network would predominantly occupy balanced configurations, while above this temperature, imbalanced configurations would become statistically significant.

\begin{figure}[t]
	\centering\includegraphics[width=0.7\linewidth]{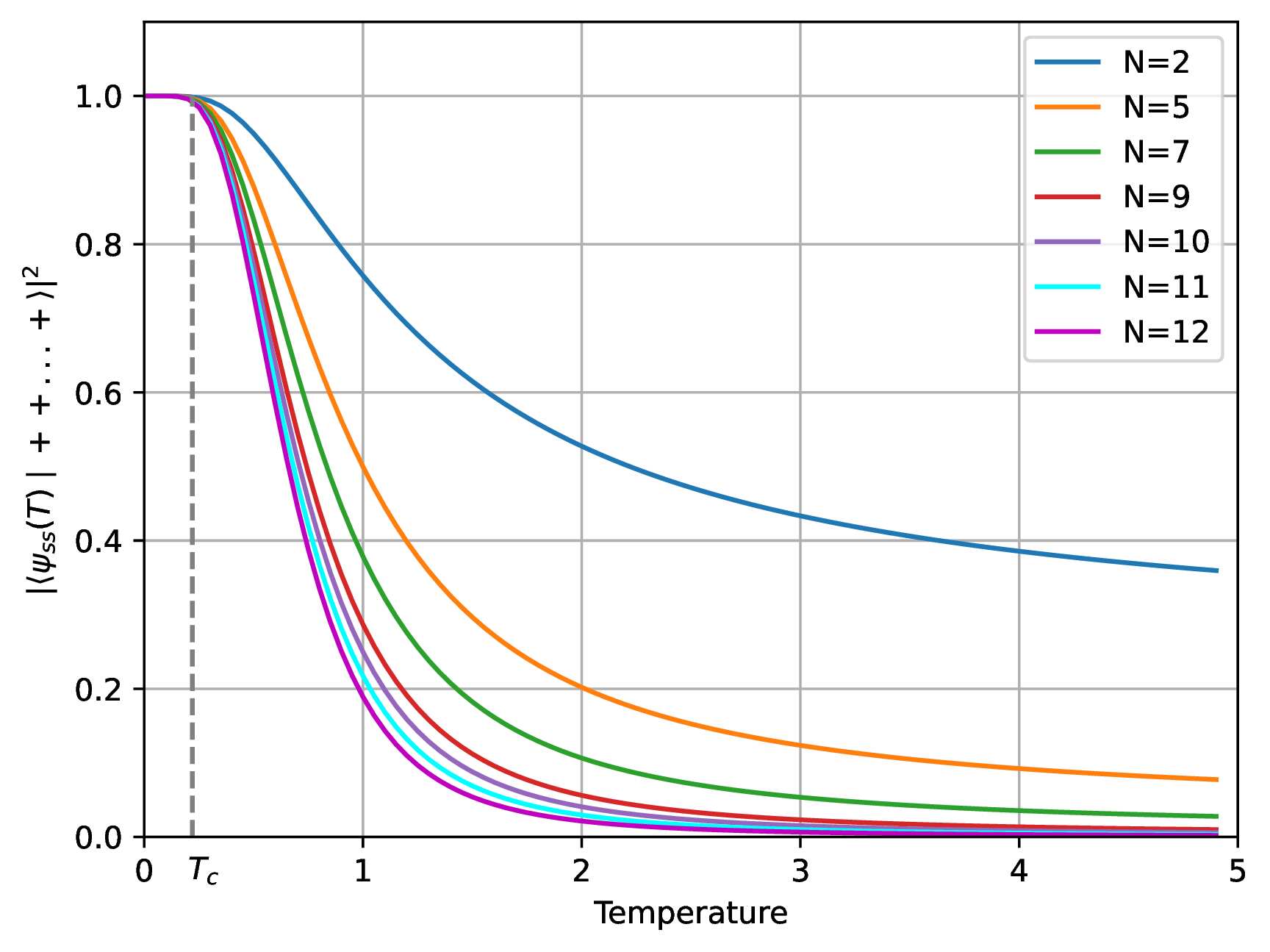}
	\caption{The diagram illustrates the system's evolution over infinite time in steady state $\ket{\psi_{ss}(T)}$ and across a range of temperatures, with the number of triads $(N)$ varying. As demonstrated in the accompanying figure, the system's evolution tends to persist in the balanced state, representing equilibrium, until the critical temperature is attained. When the temperature exceeds this value $T_c$ around $0.22$, the probability of the system being in the balanced state decreases. This decrease occurs more rapidly as the size of the system increases. }{\label{fig6}}
\end{figure}
By increasing the number of triads or qubits (N), the Hilbert space will grow with $2^N$, which poses computational challenges.  Fig. \ref{fig5}, left panel, depicts the transition dynamics at zero temperature, where transitions exclusively favor the formation of balanced triads. Under finite temperature conditions, Fig. \ref{fig6}, the system experiences two competing flows: one directed toward increased balance and another, enabled by the Boltzmann factor, toward greater imbalance. The equation gives the corresponding Hamiltonian incorporating these thermally weighted transition rates. 
\begin{align}{\label{eq_Ht}}
    H = & \sum_{k>l=1}^N  [\sigma^+_k 
    - (\frac{1}{2} \mathbb{I}_k - \hat{n}_k)] \otimes \mathbb{I}_l
    + \sum_{k>l=1}^N [\sigma^+_k \otimes \sigma^-_l
    -(\frac{1}{2} \mathbb{I}_k - \hat{n}_k) \otimes (\frac{1}{2} \mathbb{I}_l + \hat{n}_l) ]
    \nonumber \\ &+ \sum_{k>l=1}^N [\sigma^+_k \otimes \sigma^+_l  - (\frac{1}{2} \mathbb{I}_k - \hat{n}_k) \otimes (\frac{1}{2} \mathbb{I}_l - \hat{n}_l)]+ e^{-\beta}\sum_{k>l=1}^N [\sigma^-_k 
    -(\frac{1}{2} \mathbb{I}_k + \hat{n}_k) ]\otimes \mathbb{I}_l  \nonumber \\
    &  + e^{-2\beta} \sum_{k>l=1}^N [\sigma^-_k \otimes \sigma^-_l
    -(\frac{1}{2} \mathbb{I}_k + \hat{n}_k) \otimes (\frac{1}{2} \mathbb{I}_l + \hat{n}_l) ].
\end{align}

Previously, system evolution was constrained such that each transition either maintained or reduced imbalance. In the finite temperature formalism, however, transitions that increase the number of imbalanced triads become possible, introducing an "upward" flow in the energy landscape proportional to $T$. To further explore this behavior, we investigated the effect of increasing the number of triads under nonzero temperature conditions. The results of steady states $\ket{\psi(\lambda = 0)}$  for various numbers of qubits, N, are summarized in Fig.\ref{fig6}, indicating the existence of a critical temperature $T_c \approx 0.22$, beyond which imbalanced configurations become statistically prominent. Remarkably, this critical temperature remains approximately constant as the number of triads increases, suggesting a robust, size-independent transition phenomenon in the quantum balance framework. The final key distinction from classical models arises here: whereas classical transformations typically occur in discrete time steps, the quantum evolution considered here is continuous in time.

%\section*{Discussion}
\section*{Conclusion}
It is fantastic how quantum mechanics reminds us that reality is woven from countless possibilities, each waiting to be observed and nurtured. This renewal and rebirth resonate with that fundamental truth: every moment is a chance for something new to emerge from the superposition of what was and what could be.
The classical Heider balance theory is a successful theory with various applications in different disciplines. In dynamics, imbalanced triads are transformed to balance by temperature based on Boltzmann's probability. While each triad can be either balanced or imbalanced, in real life, it is our subjective experience that we are in the superposition of balanced and imbalanced triads in our relationships, with some stress and relief at the same time. Only by the events that occur may one of the states of balance or imbalance be observed.
All this became an incentive to generalize the balance theory in the framework of quantum mechanics. In this generalization, we used creation and annihilation operators to construct the Hamiltonian of the balance theory transformation from the perspective of the second quantization. Meanwhile, we learned how to rewrite the Hamiltonian of the transformations from the perspective of quantum mechanics, identifying entangled states that were not observable in classical physics. There are possible states inherent in the balance theory that fit our life experience; however, they are not observable due to the limitations of classical physics. Quantum algorithms are experiencing a phase of rapid advancement, marked by substantial progress in the exploration of possible solutions. However, the challenge of addressing computational costs remains unresolved.
\cite{2Xu2024,vondollen3,PhysRevResear4}. %To address this issue, a machine learning surrogate model inspired by quantum mechanics has been developed to improve the efficiency of optimization processes. Some research has shown the employment of random forests as a surrogate not only circumvents the necessity for costly physical modeling but also expedites convergence while preserving accuracy \cite{2Xu2024,vondollen3,PhysRevResear4}.

\section*{Acknowledgment}
The authors gratefully acknowledge Abolfazl HaqiqiFar, and Mohammad Amin Safaei for reading the paper and providing comments.

\bibliographystyle{unsrt}  

\bibliography{template}

\end{document}